\documentclass[preprint,preprintnumbers,amsmath,amssymb]{revtex4}

\usepackage{graphicx}
\usepackage{dcolumn}
\usepackage{bm}

\begin{document}

\preprint{The following article was submitted to the \textit{Review of Scientific Instruments}.}

\title{Apparatus for real-time acoustic imaging of Rayleigh-B\`{e}nard convection}
\author{Kerry Kuehn}
\homepage{http://www.faculty.wlc.edu/kuehn}
\email{kerry.kuehn@wlc.edu}
\author{Jonathan Polfer}
 \altaffiliation[Currently at ]{Abbott Laboratories, Abbot Park, Illinois.}
\author{Joanna Furno}
\author{Nathan Finke}
\affiliation{Wisconsin Lutheran College, Milwaukee, WI 53226}

\date{\today}

\begin{abstract}
We have designed and built an apparatus for real-time acoustic imaging of convective flow patterns in optically opaque fluids.  This apparatus takes advantage of recent advances in two-dimensional ultrasound transducer array technology; it employs a modified version of a commercially available ultrasound camera, similar to those employed in non-destructive testing of solids.  Images of convection patterns are generated by observing the lateral variation of the temperature dependent speed of sound \textit{via} refraction of acoustic plane waves passing vertically through the fluid layer.  The apparatus has been validated by observing convection rolls in both silicone oil and ferrofluid.   
\end{abstract}

\maketitle

\section{\label{sec:intro}Introduction}

Thermal convection has been studied intensely in the laboratory for years \cite{Rumford:1965ij}.  A relatively simple arrangement which is commonly used to study thermal convection is a thin horizontal layer of fluid confined between flat horizontal plates subjected to a vertical temperature gradient.  In this arrangement, when the temperature of the bottom plate exceeds that of the top plate by a critical value, the fluid layer becomes mechanically unstable.  This, the so called Rayleigh--B\`{e}nard instability  \citep{Benard:1900fk, Rayleigh:1916fk}, is identified by a marked increase in the thermal conductivity of the fluid layer due to the onset of Rayleigh--B\`{e}nard convection  \citep{schmidt:m:1935}.

Shadowgraph visualization of the Rayleigh-B\`{e}nard instability in optically transparent fluids has enabled comparison between theoretical and experimental work on a well defined nonlinear system.  Despite the simplicity of this arrangement and knowledge of the underlying dynamical equations of fluid motion, the nature of the sequence of transitions from diffusive to time--dependent and finally to turbulent heat transport in this system is not understood \citep{bodenschatz:p:a:2000}.  The weakly non--linear theory of convection is able to predict the critical Rayleigh number, the critical wavenumber, and the pattern selected at the onset of convection \citep{schluter:l:b:1965}.  When the critical temperature difference across the fluid layer is relatively small, such that the fluid properties vary little within the fluid layer, one expects to observe a continuous transition from a uniform base state to a pattern consisting of stationary two--dimensional convection rolls.  Such patterns have been observed in simple fluids at onset in both rectangular and axisymmetric convection cells \citep{koschmieder:1966}.  The weakly non--linear theory is incapable, however, of describing moderately supercritical convection.  

The non-linear properties of moderately supercritical convection have been studied in detail by Busse and collaborators \citep{busse:1978}.  By numerically solving the non--linear Oberbeck--Boussinesq equations of fluid motion for a laterally unbounded fluid layer confined between horizontal plates in the presence of a vertical thermal gradient, they have been able to construct a stability diagram, known as the ``Busse balloon,'' for convective flow covering a wide range of Rayleigh ($R\!a$) and Prandtl ($P\!r$) numbers.
\begin{equation}
  R\!a = \frac{ \alpha g d^3 \Delta T }{ \nu \kappa }~~~~~~~~~~~P\!r = \frac{\nu}{\kappa}
  \label{eq:Rasigma} \nonumber
\end{equation} 
\noindent The Rayleigh number serves as a control parameter for the onset of convection.  The Prandtl number dictates the nature of the secondary instabilities to which convection rolls are subject above onset.  In these equations, $d$ is the horizontal plate separation, $\Delta T$ is the temperature difference between the horizontal plates, $g$ is the gravitational field strength, and $\alpha$, $\nu$ and $\kappa$ are the isobaric thermal expansion coefficient, the kinematic viscosity and the thermal diffusivity of the fluid.  

Early studies using helium \citep{ahlers:1974kx} revealed a time--dependent Nusselt number just above onset.  The Nusselt number, $N\!u$, is just the dimensionless ratio of the observed thermal conductivity to that of the quiescent fluid.  Such a time dependent response to a constant driving force suggests a time--dependent flow pattern.  Unfortunately, flow visualization was not available for these experiments, and measurements of the Nusselt number alone only provide information on the global, i.e.\ the spatially averaged, character of heat transport.  Direct visualization of the fluid, on the other hand, allows features in the global heat transport to be identified with specific dynamic events and spatial structures.  In particular, shadowgraph visualization from above the convection cell has been used extensively to obtain flow images which have been indispensable for comparisons of theoretical calculations and experimental measurements in optically transparent fluids \citep{chen:w:1968, pocheau:c:l:1985, debruyn:b:m:t:h:c:a:1996}.

A similar comparison between theory and experiment in many opaque fluids has not been made, owing primarily to the difficulty of imaging flow patterns in optically opaque fluids.  For example, of the extant studies of Rayleigh-B\`{e}nard convection in liquid metals, only Fauve, Laroche, and Libchaber have visualized convective flow from above \citep{fauve:l:l:1981}.  They used a temperature-sensitive liquid crystal sheet beneath a transparent (sapphire) top plate to visualize localized temperature differences associated with convective structures in liquid mercury confined to small aspect ratio cells.   These studies, however, lacked detailed images of convective flow patterns.  There has been no systematic experimental study of the convective planform for Rayleigh-B\`{e}nard convection in a liquid metal.  This is troubling, since the Prandtl numbers of liquid metals are typically an order of magnitude (or two, or three) lower than that of other fluids, and since the Prandtl number is expected to play an important role in determining the nature of the instabilities to which straight convective rolls are subject above the onset of convection \citep{busse:1978}.  

The liquid crystal method has proved rather more fruitful in the study of ferrofluid convection.  Bozhko and Putin \citep{Bozhko:2003fk}, for instance, have used this method to visualize convection patterns in a kerosene-based magnetic fluid confined to a cylindrical cell.  In the absence of an applied magnetic field, they observed oscillatory convection for Raleigh numbers up to $4 R\!a_c$.   On the other hand, Huke and L\"ucke have calculated the convection patterns for typical ferrofluid parameters, and found that squares should be the first stable pattern to appear above onset \citep{Huke:2005fk}.  For ferrofluids, like other binary mixtures, the nature of the instabilities to which the squares are subject depends upon other fluid parameters such as the separation ratio and the Lewis number.

Generally speaking, the liquid crystal method is limited by the thermal response of the liquid crystal.  In the experiments of Bozhko and Putin, for instance, the entire color change of the liquid crystal occurs between 24 and 27 degrees Celsius.  Moreover, the liquid crystal must be thermally isolated from the sample fluid in order to protect the liquid crystal from the chemical action of the ferrofluid.  As an alternative to the liquid crystal method, the use of ultrasound has been explored for use in acoustic imaging of Rayleigh-B\'{e}nard convection.  Most recently, Xu and Andereck \citep{Xu:2003kx} used an array of 11 contact transducers fixed to the side of a small aspect ratio convection cell to obtain a two dimensional map of the thermal field in liquid mercury undergoing convection.  Their method, however, does not allow one to study the convective planform from above, as one can with standard shadowgraph and liquid crystal techniques.

In the present paper, we describe an apparatus which allows real-time acoustic imaging of thermal convection in a Rayleigh-B\`{e}nard cell.  This apparatus can be used to visualize thermal convection, from above the convection cell, in optically opaque fluids such as liquid metals, ferrofluids and opaque gels.  The paper is organized as follows.  In Sec.~\ref{sec:design}, we describe the design and construction of the apparatus.  Here, we provide an overview of the apparatus, followed by detailed descriptions of the ultrasound imaging system, the convection cell, thermometer construction and temperature regulation.  In Sec.~\ref{sec:experiments}, we describe experimental procedures and results.  In particular, we provide images of convective structures observed in silicone oil and ferrofluid.  We also describe prospects for continuing studies of ferrofluids and for planned work using liquid mercury.

\section{\label{sec:design}Apparatus design and construction}

\subsection{\label{subsec:overview}Overview}

A sectional view of the apparatus design is shown in Fig.~\ref{fig01}.  The sample fluid is introduced \textit{via} one of two entrance ports ($A$) to a sample space located between two sapphire windows ($B$).  These windows form the top and bottom boundaries of the convection cell and are separated by a thin spacing ring ($C$).  A detailed description of the convection cell is provided in Sec.~\ref{subsec:cell}.

A vertical temperature gradient is maintained across the convection cell by thermally regulating the water circulating through the spaces above the top, and below the bottom, sapphire windows.  First,  we describe the thermal regulation of the top sapphire window.  Cool water from a refrigerated bath circulator enters the apparatus through a port ($D$) and flows through a channel which runs around the circumference of the  apparatus ($E$).  The chilled water is forced downward through a series of six vertical holes (not shown) in the floor of this channel and into a second flow distribution channel ($F$).  

From the flow distribution channel, the water is injected into the top bath space through six jets ($G$) spaced evenly about the perimeter of the bath space.  These jets direct the chilled water onto the top surface of the upper sapphire window.  Heat which has propagated vertically through the cell is absorbed by the chilled water.  The chilled water leaves the bath space through six exit ports ($H$), which are spaced evenly about the top of the bath space.  After flowing through another circumferential channel ($I$), the fluid returns to the bath circulator \textit{via} the exit port ($J$).  

Thermal regulation of the bottom sapphire window is accomplished in a similar fashion by circulating warm water through the lower bath space.  The temperature of the upper and lower bath spaces are monitored using thermometers inserted through water-tight radial ports ($K$) in the wall of the upper and lower bath spaces.  Up to four thermometers may be inserted into each bath space.  A detailed description of the thermometer construction is provided in Sec.~\ref{subsec:thermometry}.  

The primary difference between the upper and the lower bath space construction is that the ultrasound transducer ($L$) is situated just below the bottom sapphire window in the lower bath space.  It is kinematically mounted atop three stainless steel threaded posts ($M$) which are inserted through bushings in the bottom bath flange ($N$).  The transducer is pulled downward against the top of the three mounting posts using a small spring (not shown) which is stretched between the bottom of the transducer and the bottom bath flange.  There are 80 threads per inch on the mounting posts, and by adjusting them appropriately, the transducer can be placed very near to the bottom sapphire window and leveled optimally.  

\subsection{\label{subsec:transducer}Ultrasound imaging system}

An exploded view of the lower bath space and the transducer is shown in Fig.~\ref{fig02}.   An RG-174 coaxial cable (not shown) which carries the ultrasound signal to the transducer is fed through a water-tight compression seal in a small hole near one of the mounting posts in the bottom bath flange.  During an experiment, the transducer generates collimated ultrasonic plane waves which propagate vertically through the sample cell and the upper bath space. These ultrasound waves are detected by a modified version of a commercially available ultrasound camera (Acoustocam Model I400, Imperium Inc., Silver Spring, MD) mounted atop an acrylic window (labeled $O$ in Fig.~\ref{fig01}) which seals the upper bath space.  Images of convection patterns within the convection cell are generated by observing the lateral variation of the temperature dependent speed of sound \textit{via} refraction of acoustic plane waves passing vertically through the sample fluid layer.

A photograph of the ultrasound camera (turned lens-side up) is shown in Fig.~\ref{fig03}.   Labeled are the acoustic lenses ($A$) which focus the ultrasound waves onto a small piezoelectric transducer array inside the body of the camera.  The array measures 12 mm on a side and consists of $120 \times 120$ individual electrodes with 100--micron center--to--center spacing.  Voltage variations in each pixel induced by incident ultrasound waves are read out using a multiplexer and frame--grabber electronics.  The image capture rate is 30 frames per second, similar to that of a standard CCD camera. The camera may be focused and the field of view may be varied by moving the acoustic lens using a knurled knob ($B$) on the side of the camera.  A field of view as large as 1-2 inches has been achieved.  The acoustic lenses are enclosed in a cylindrical shield (not shown) which is filled with an impedance matching fluid \textit{via} an external port ($C$).  

\subsection{\label{subsec:control}External connections}

Once the apparatus is assembled, it is placed on a tripod mount atop an aluminum platform, as shown in the photograph of Fig.~\ref{fig04}.  A shallow tray placed under the platform catches any spillage in the event of a leak.  The black tubes in the figure are thermally insulated tygon tubes which carry cooling and heating water to and from the bath spaces.  The white wires in the figure are electrically insulated leads for the thermometers.  Also shown is a thin tygon tube which is attached to a syringe and which is used to fill the cell with a sample fluid.  A vertical instrumentation panel on the front of the aluminum platform is fitted with feed-throughs to facilitate electrical and fluid connections between the apparatus and the external devices such as the sample fluid syringe, the bath circulators, the data acquisition/switch unit and the ultrasound controller.   

The external connections to the apparatus are depicted schematically in Fig.~\ref{fig05}.  They can be divided into three separate systems:  bath control, temperature monitoring, and ultrasound imaging.  These systems are controlled using software running on a PC, as described in the next paragraph. 

The top and bottom bath space temperatures are regulated using separate refrigerated bath circulators (NESLAB RTE-7 Digital Plus Refrigerated Bath, Thermo Fisher Scientific, Waltham, MA).  Set-point temperatures are assigned from software by sending RS-232 commands \textit{via} a cable connection between the PC and the bath circulators.  The temperature of the upper and lower bath spaces inside the apparatus are monitored using six thermometers.  Temperature values of the six thermometers are scanned twice per second using a digital data acquisiton unit (Model 2700 DMM, Keithley Instruments, Cleveland, OH) connected to a USB port on the PC using a USB-to-GPIB interface adapter (Model KUSB-488A, Keithley Instruments, Cleveland, OH).  The ultrasound imaging system is controlled using a commercial software package (AcoustoVision400 software, Imperium, Inc., Silver Springs, MD).  The software allows the user to communicate with an ultrasound controller box which in turn attaches to both the transducer and the camera.  

\subsection{\label{subsec:focus}Image calibration and resolution}

The ultrasound images were calibrated by placing an object of known dimensions in the sample space and recording the number of image pixels corresponding to its size.  For this purpose we fabricated an array of small holes drilled at precisely measured intervals into a thin sheet of aluminum.  We placed this ``phantom" between the sapphire windows of the convection cell in order to calibrate our images.  

We also placed some more mundane objects into the cell so as to illustrate the resolution limit which is imposed by diffraction effects.  Shown in Fig.~\ref{fig06} are acoustic images of an ordinary paper clip and a small stainless steel washer which were placed on top of the convection cell.  The successive images, (a) through (d), illustrate the observations as we progressively moved the acoustic lenses so as to focus on the objects.  The small ripples which can be observed are artifacts caused by diffraction.  The resolution of the camera is diffraction limited by the wavelength of the ultrasound, which in turn depends upon the speed of sound in the medium.  The wavelength of 5 MHz ultrasound in water is about 0.3 mm.        

Another imaging difficulty which was encountered involved the accumulation of bubbles on the top bath window ($O$ in Fig.~\ref{fig01}) due to outgassing of the circulating bath water.  These bubbles seriously degraded the ultrasound image quality during the course of consecutive experiments.  To resolve this issue, the bath water was typically degassed before being added to the circulator.  This, however, did not solve the problem, since over time the circulating water would eventually reabsorb air.  The final solution was to design a small wiper which could be swept across the surface of the bath window in-situ so as to clear the bubbles from the ultrasound path during the course of an experiment, if necessary.  

The bubble wiper consisted of a plastic rectangular prism 1/8 inch $\times$ 1/8 inch $\times$ 2.75 inches.  A strip of doll eyelashes (Michaels, West Allis, WI) was attached to the long edge of the prism with water-resistant glue.  A thin stainless steel guide wire was slid through a small hole drilled in the long edge of the prism and secured in place with a set screw.  In this way, the bubble wiper resembled a tiny broom or rake, with the guide wire the handle.  The guide wire was fed through a stainless steel capillary tube which was itself slid through a water-tight seal in the upper edge of the bath space just below the top bath window.  Pushing on the guide wire caused the bubble wiper to slide along a set of tracks so that the doll hairs brushed along the top bath window.  In this way, the bubbles were pushed to one edge of the window.  After pushing the bubbles aside, they were removed from the bath space through a tiny capillary tube which was inserted through another tiny hole in the wall of the bath.  The tip of the capillary tube resided near the top bath window, and by using a syringe, the accumulated air bubbles could be easily removed from the bath space.  This contraption proved very effective in clearing bubbles from the field of view and permitting clear ultrasound images to be obtained. 

\subsection{\label{subsec:cell}Convection cell details}

The windows which form the horizontal boundaries of the convection cell are four inch diameter and $0.2$ inch thick optically flat single-crystal sapphire disks (Hemlite sapphire windows, Crystal Systems Inc., Salem, MA).  Sapphire was chosen because of its high thermal conductivity and its small ultrasonic attenuation coefficient.  The thermal conductivity of sapphire is approximately 50 times greater than that of quiescent liquid mercury, the sample fluid with the highest thermal conductivity which we plan to study.  A high relative thermal conductivity minimizes undesirable horizontal thermal gradients in the windows.

A thin spacing ring establishes the height, $d$, of the convection cell.  Recall that the critical temperature difference, $\Delta T_c$, for the Rayleigh--B\`{e}nard instability is inversely proportional to $d^3$.  $\Delta T_c$, in turn, establishes the minimum heating and cooling power of the bath circulators:  a shallow fluid layer requires a large temperature difference;  a deep fluid layer requires a conveniently small temperature difference, but increases the aspect ratio of the cell, which is undesirable.  The aspect ratio of the convection cell, $\Gamma$, is defined as the dimensionless ratio of the cell width to its depth.  Both the critical Rayleigh number and the convective planform are generally influenced by the existence of lateral sidewalls \citep{koschmieder:1993}.  If one wishes to study the stability of the fluid layer itself, sidewall effects are considered a spurious complication.  

In order to minimize the effects of the experimentally unavoidable lateral sidewalls, it is desirable to employ a large aspect ratio cylindrical cell, since this has the smallest circumference of any shape of a given cross--sectional area.  Moreover, the thermal conductivity of the spacing ring should ideally match that of the fluid so as to prevent lateral thermal gradients near the sidewall.  Hence the material and dimensions of the spacing ring depends upon the choice of fluid being studied.  

We constructed cells with five different spacing ring materials and dimensions.  The properties of these cells are summarized in Tab.~\ref{tab:cells}.  The first, and tallest, cell is designed specifically to study liquid mercury.  Stainless steel was chosen as the spacing ring material because its thermal conductivity ($1.6 \times 10^5$ erg/s$\cdot$cm$\cdot$K) is only slightly greater than that of liquid mercury ($8.3 \times 10^4$ erg/s$\cdot$cm$\cdot$K).  Since above onset the Nusselt number increases with Rayleigh number, the thermal conductivity of convecting mercury increasingly matches that of stainless steel until it reaches approximately 30,000, which lies significantly above the expected stability region for straight rolls.  The remaining four cells were constructed with acetal sidewalls and have different heights so as to allow the onset of convection to occur at different values of $\Delta T$.  

After assembling the cell and before assembling the rests of the apparatus, it is important to verify that the sapphire windows are parallel.  This was done using two techniques.  The first was by using a micrometer.  The spacing from the top of the top sapphire window to the bottom of the bottom sapphire window was measured at several locations across the cell.  The second method was by using an interferometer setup, as shown in Fig.~\ref{fig07}.  A He-Ne laser beam was expanded and collimated to a beam width of approximately one inch.  A pellicle beam splitter (Melles Griot, Rochester, NY) was used to direct the collimated beam at the cell and also at a screen, on which appeared interference fringes when the cell windows were not perfectly aligned.  Typically, the interference fringes were concentric rings, less than ten in number, and clustered near the outside perimeter of the cell.  This implies that the central region of the cell is quite parallel, and that the lack of parallelism is typically less than about 30 microns and occurs near the cell edges.  In the event that the windows are not aligned properly, the screws which hold the cell together may be adjusted so as to establish optimal parallelism.   

\subsection{\label{subsec:thermometry}Thermometry details}

The temperature near the sapphire windows was monitored using thermometers constructed from glass encapsulated thermistors.  One such thermometer is shown in Fig.~\ref{fig08}.  A thermistor hermetically sealed in a thin glass rod (P60 NTC bead-in-glass thermoprobe, P.N. P60DB303M, GE Thermometrics, Edison, NY) was inserted into a 1/16 inch hole in a nylon female luer bulkhead adapter (P.N. B-1005126, Small Parts, Inc., Miami Lakes, FL).  Waterproof epoxy (Loctite Fixmaster High Performance Epoxy, P.N.~99393, Henkel Corporation, Rocky Hill, CT) was used to seal the thermistor to the luer connector.  The two leads of the thermistor were electrically insulated with a short length of 0.025 inch diameter spaghetti tubing (P.N. ETT-30, Weico, Edgewood, NY).  A 40 centimeter long twisted pair of manganin wires (MW 36-AWG, Lakeshore Cryogenics, Westerville, OH) was then soldered to the tip of each of the leads of the thermistor.  This configuration allowed for four-wire resistance measurements.  The solder joints were insulated with heat-shrink tubing and each twisted pair was inserted into a length of teflon tubing (PTFE 20 TW, McMaster-Carr, Chicago, IL).  Heat shrink tubing was wrapped around the joint between the luer connector and the wires for added strain relief.  Finally, pin connectors were soldered to the end of the wires.

The assembled thermometers are screwed into threaded ports in the sidewall of the top and bottom bath spaces.   A tiny o-ring establishes a water-tight seal.  The tip of each thermistor projects about a half an inch into the bath space and resides approximately 1/16 of an inch from the outside surface of the sapphire windows.  During an experiment, the bath temperatures are sampled twice per second.  Shown in the top graph of Fig.~\ref{fig09} is the temperature versus time of one of the thermometers during a typical experiment.  In this particular case, the top and bottom bath spaces were both regulated near 20 degrees Celsius.  The root mean square temperature noise is less than 10 mK.  Shown in the bottom graph of Fig.~\ref{fig09} is the power spectral density computed from this same data.   

\section{\label{sec:experiments}Experiments}

\subsection{\label{subsec:procedure}Procedure}

The following procedure is typically followed in performing a set of experiments.  First, the convection cell is assembled with the sidewall appropriate for the sample fluid to be studied.  The convection cell is tested for leaks and the sapphire windows are tested for parallelism.  The remainder of the apparatus is assembled around the cell and the apparatus is placed on the tripod on the aluminum platform.  Before screwing on the top window ($O$ in Fig.~\ref{fig01}), a small bubble level is placed on the top sapphire window to ensure that the apparatus is level.  Next, the bath circulators are filled with degassed water mixed with algicide (Chloramine-T, Thermo Fisher Scientific, Waltham, MA).  

The bath circulator tubing is then attached to the apparatus and the top and bottom bath spaces are filled with water.  While the bath spaces are being filled, nitrogen gas is circulated through the convection cell to keep it clean.  At this point, the thermometers are plugged in to the data acquisition unit and the temperature monitoring software is started.    

Next, a syringe is used to push the sample fluid into the convection cell.  Typically, the apparatus is tipped onto its side, with the cell entrance port ($A$ in Fig.~\ref{fig01}) on the bottom, while filling the convection cell.  The cell is seen to be full when sample fluid squirts out of the exit port on the top, opposite the entrance port.  The apparatus is returned to its upright position on the aluminum platform and it is visibly inspected for leaks.  The ultrasound camera is then attached to the top of the apparatus and is plugged into the ultrasound controller box.  The image capture software is started and the camera settings are adjusted for a clear ultrasound image.

In a typical set of experiments, the temperature difference across the cell, $\Delta T$, is initially maintained at close to zero degrees.  Several background images of the quiescent fluid are captured.  Next, $\Delta T$ is increased stepwise, maintaining the average fluid temperature constant.  Since the Prandtl number is known to be weakly temperature dependent, this technique serves to keep the mean Prandtl number of the fluid constant.  At each step, a series of ultrasound images is obtained.  The system is allowed to equilibrate for a minimum of one horizontal diffusion time after each step.  The horizontal diffusion time is given by $\tau_h =\Gamma \tau_v$ and the vertical diffusion time is given by $\tau_v = d^2/\kappa$ (see Tab.~\ref{tab:properties}).

After data collection, the images are analyzed on a personal computer using image processing software (Igor Pro v.5, WaveMetrics, Inc., Lake Oswego, OR).  The onset of convection is determined by the appearance of spatial variations in the temperature dependent acoustic refractive index.  Such spatial variations are analyzed so as to determine the convective planform and its wavenumber distribution.  

\subsection{\label{subsec:results}Results and Discussion}

We present here the first acoustic images of the planform of Rayleigh-B\'{e}nard convection viewed from above.  Shown in Fig.~\ref{fig10} is the onset of convection in 5 cSt polydimethylsiloxane (PDMS) polymer fluid (Dow Corning 200 fluid, Dow Corning, Midland, MI), otherwise known as silicone oil.  Part (a) of the figure depicts the uniform conductive state, below the onset of convection.  Here,  $\Delta T = 3.22$ degrees and $R\!a = 910$.  Just below, in part (b) of the figure, is shown the convective state.  Here, $\Delta T = 10.89$ degrees and $R\!a = 3082$.  In both cases, the average fluid temperature was 28.5 degrees and the Prandtl number was 67.8.  The short tick marks along the horizontal and vertical axes indicate lateral distances in units of the cell height, $d$.  The images have been normalized by (i) subtracting a background image obtained at very low Rayleigh number and then (ii) dividing the resulting image by the same background image.  

To the immediate right of images (a) and (b) are shown their respective power spectral densities.  These were obtained by the following procedure.  First, a Kaiser window was applied to the real-space image.  Next, the two-dimensional fourier transform of the image was computed.  A high-pass filter was applied to the real and imaginary parts of the transformed image to filter out the very low frequency background components.  The one-sided power spectral density was then computed from the sum of the squares of the real and imaginary parts, and this power spectral density was normalized so as to satisfy Parseval's theorem.  

The axes of the computed power spectral density indicate the cartesian components of the wavenumber, $k$.  The wavenumber is defined as the ratio of the spatial wavelength, $\lambda$, and the cell height, $d$.  The lack of structure in the power spectral density of image (a) reflects a uniform conductive state.  The (black) peak in the power spectral density of image (b) reflects the appearance of straight convection rolls at the indicated value of $k$.  Investigation of the variation of the power spectral density with Rayleigh number provides a quantitative method for studying the onset and the evolution of Rayleigh-B\'{e}nard convection.   

Shown in Fig.~\ref{fig11} are acoustic images of two convective states observed in EFH1 ferrofluid (P.N. FF-310, Educational Innovations, Inc., Norwalk, CT).  This opaque fluid is a colloidal dispersion comprised of magnetite particles (3-15\% by volume) and oil soluble dispersant (6-30\% by volume) suspended in a carrier liquid (55-91\% by volume).  For this particular set of experiments, the convection cell was tilted by 2.89 degrees so as to induce a large scale lateral flow and to facilitate roll formation.  Part (a) of the figure depicts a state of cellular convection just above onset.  Here,  $\Delta T = 15.44$ degrees and $R\!a = 1986$.  Part (b) of the figure depicts a straight-roll convection state at somewhat higher Rayleigh number.  Here, $\Delta T = 26.89$ degrees and $R\!a = 3460$.  For these experiments, the average fluid temperature was 29.9 degrees  and the Prandtl number was 101.2.  As in Fig.~\ref{fig10}, to the immediate right of each image is shown a grayscale image of its computed power spectral density.  The broad peak near $k =1$ in the power spectral density associated with image (a) reflects the lack of orientational order observed in the cellular convection state, whereas the localized peak in the power spectral density associated with image (b) reflects the distinct straight roll pattern.

As mentioned in Sec.~\ref{sec:intro}, there exist detailed theoretical predictions for the stability diagram of binary fluids such as ferrofluids.  A systematic experimental study of the variation of the power spectral density as a function of Rayleigh number in ferrofluids using acoustic imaging is being prepared for a future publication.  In addition to work on ferrofluids, future experiments will focus on the onset of Rayleigh-B\`{e}nard convection in liquid mercury.  Some of the properties of liquid mercury are presented in Tab.~\ref{tab:properties}.  Liquid mercury is a good choice from among the low Prandtl number liquid metals due to its relatively large thermal expansion coefficient (witness the mercury thermometer) and hence its propensity to exhibit convection.  Based on the theoretical stability diagram for liquid mercury  \citep{busse:1978}, we anticipate that straight rolls should be stable over a range of Rayleigh numbers, between 1708 and approximately 1900.  Thus, in our cell number 1, whose height is 0.514 centimeter, we anticipate straight rolls to be stable when the temperature difference across the cell is between 4.82 and 5.37 degrees Celsius, giving a stability range of approximately 0.55 degrees Celsius.  For shorter cells, the stability range is larger.  This tendency is summarized in Tab.~\ref{tab:hg}, where we list the anticipated stability ranges for straight rolls in liquid mercury for each of our five convection cells.

\begin{acknowledgments}
Thanks to Jack Gurney, formerly at Imperium, Inc., for providing extensive engineering consultation and support during the initial stages of the project.  Thanks to Richard Scherr in the biophysics machine shop of the Medical College of Wisconsin for helping to build a preliminary version of our convection cell.  Finally, thanks to Professor Daniel Ebeling in the Chemistry Department at Wisconsin Lutheran College and Cindy Kuehn for helpful suggestions.  

This work was supported by a Major Research Instrumentation grant from the National Science Foundation (DMR-0416787) and a grant from the Office of Basic Energy Sciences at the U.S. Department of Energy (DE-FG02-04ER46166). 
\end{acknowledgments}


\begin{thebibliography}{21}
\expandafter\ifx\csname natexlab\endcsname\relax\def\natexlab#1{#1}\fi
\expandafter\ifx\csname bibnamefont\endcsname\relax
  \def\bibnamefont#1{#1}\fi
\expandafter\ifx\csname bibfnamefont\endcsname\relax
  \def\bibfnamefont#1{#1}\fi
\expandafter\ifx\csname citenamefont\endcsname\relax
  \def\citenamefont#1{#1}\fi
\expandafter\ifx\csname url\endcsname\relax
  \def\url#1{\texttt{#1}}\fi
\expandafter\ifx\csname urlprefix\endcsname\relax\def\urlprefix{URL }\fi
\providecommand{\bibinfo}[2]{#2}
\providecommand{\eprint}[2][]{\url{#2}}

\bibitem[{\citenamefont{Rumford}(1965)}]{Rumford:1965ij}
\bibinfo{author}{\bibfnamefont{C.}~\bibnamefont{Rumford}}, in
  \emph{\bibinfo{booktitle}{Source Book in Physics}}, edited by
  \bibinfo{editor}{\bibfnamefont{E.~H.} \bibnamefont{Madden}}
  (\bibinfo{publisher}{Harvard University Press}, \bibinfo{address}{Cambridge,
  Massachusetts}, \bibinfo{year}{1965}), pp. \bibinfo{pages}{146--161},
  \bibinfo{note}{read before the Royal Society in 1798}.

\bibitem[{\citenamefont{{Benard}}(1900)}]{Benard:1900fk}
\bibinfo{author}{\bibfnamefont{H.}~\bibnamefont{{Benard}}},
  \bibinfo{journal}{{Rev. Gen. Sci. Pures Appl}}  (\bibinfo{year}{1900}).

\bibitem[{\citenamefont{{Rayleigh}}(1916)}]{Rayleigh:1916fk}
\bibinfo{author}{\bibfnamefont{L.}~\bibnamefont{{Rayleigh}}},
  \bibinfo{journal}{Philosophical Magazine and Journal of Science}
  \textbf{\bibinfo{volume}{32}}, \bibinfo{pages}{529} (\bibinfo{year}{1916}).

\bibitem[{\citenamefont{Schmidt and Milverton}(1935)}]{schmidt:m:1935}
\bibinfo{author}{\bibfnamefont{R.~J.} \bibnamefont{Schmidt}} \bibnamefont{and}
  \bibinfo{author}{\bibfnamefont{S.~W.} \bibnamefont{Milverton}},
  \bibinfo{journal}{Proc. Roy. Soc. London A} \textbf{\bibinfo{volume}{152}},
  \bibinfo{pages}{586} (\bibinfo{year}{1935}).

\bibitem[{\citenamefont{Bodenschatz et~al.}(2000)\citenamefont{Bodenschatz,
  Pesch, and Ahlers}}]{bodenschatz:p:a:2000}
\bibinfo{author}{\bibfnamefont{E.}~\bibnamefont{Bodenschatz}},
  \bibinfo{author}{\bibfnamefont{W.}~\bibnamefont{Pesch}}, \bibnamefont{and}
  \bibinfo{author}{\bibfnamefont{G.}~\bibnamefont{Ahlers}},
  \bibinfo{journal}{Annu. Rev. Fluid Mech.} \textbf{\bibinfo{volume}{32}},
  \bibinfo{pages}{709} (\bibinfo{year}{2000}).

\bibitem[{\citenamefont{Schl\"{u}ter et~al.}(1965)\citenamefont{Schl\"{u}ter,
  Lortz, and Busse}}]{schluter:l:b:1965}
\bibinfo{author}{\bibfnamefont{A.}~\bibnamefont{Schl\"{u}ter}},
  \bibinfo{author}{\bibfnamefont{D.}~\bibnamefont{Lortz}}, \bibnamefont{and}
  \bibinfo{author}{\bibfnamefont{F.}~\bibnamefont{Busse}}, \bibinfo{journal}{J.
  Fluid Mech.} \textbf{\bibinfo{volume}{23}}, \bibinfo{pages}{129}
  (\bibinfo{year}{1965}).

\bibitem[{\citenamefont{Koschmieder}(1966)}]{koschmieder:1966}
\bibinfo{author}{\bibfnamefont{E.~L.} \bibnamefont{Koschmieder}},
  \bibinfo{journal}{Beitr. Phys. Atmos.} \textbf{\bibinfo{volume}{39}},
  \bibinfo{pages}{1} (\bibinfo{year}{1966}).

\bibitem[{\citenamefont{Busse}(1978)}]{busse:1978}
\bibinfo{author}{\bibfnamefont{F.~H.} \bibnamefont{Busse}},
  \bibinfo{journal}{Rep. Prog. Phys.} \textbf{\bibinfo{volume}{41}},
  \bibinfo{pages}{1929} (\bibinfo{year}{1978}).

\bibitem[{\citenamefont{{Ahlers}}(1974)}]{ahlers:1974kx}
\bibinfo{author}{\bibfnamefont{G.}~\bibnamefont{{Ahlers}}},
  \bibinfo{journal}{Physical Review Letters} \textbf{\bibinfo{volume}{33}},
  \bibinfo{pages}{1185} (\bibinfo{year}{1974}).

\bibitem[{\citenamefont{Chen and Whitehead}(1968)}]{chen:w:1968}
\bibinfo{author}{\bibfnamefont{M.~M.} \bibnamefont{Chen}} \bibnamefont{and}
  \bibinfo{author}{\bibfnamefont{J.~A.} \bibnamefont{Whitehead}},
  \bibinfo{journal}{J. Fluid Mech.} p.~\bibinfo{pages}{1}
  (\bibinfo{year}{1968}).

\bibitem[{\citenamefont{Pocheau et~al.}(1985)\citenamefont{Pocheau, Croquette,
  and L{e G}al}}]{pocheau:c:l:1985}
\bibinfo{author}{\bibfnamefont{A.}~\bibnamefont{Pocheau}},
  \bibinfo{author}{\bibfnamefont{V.}~\bibnamefont{Croquette}},
  \bibnamefont{and} \bibinfo{author}{\bibfnamefont{P.}~\bibnamefont{L{e G}al}},
  \bibinfo{journal}{Phys. Rev. Lett.} \textbf{\bibinfo{volume}{55}},
  \bibinfo{pages}{1094} (\bibinfo{year}{1985}).

\bibitem[{\citenamefont{deBruyn et~al.}(1996)\citenamefont{deBruyn,
  Bodenschatz, Morris, Trainoff, Hu, Cannell, and
  Ahlers}}]{debruyn:b:m:t:h:c:a:1996}
\bibinfo{author}{\bibfnamefont{J.~R.} \bibnamefont{deBruyn}},
  \bibinfo{author}{\bibfnamefont{E.}~\bibnamefont{Bodenschatz}},
  \bibinfo{author}{\bibfnamefont{S.~W.} \bibnamefont{Morris}},
  \bibinfo{author}{\bibfnamefont{S.~P.} \bibnamefont{Trainoff}},
  \bibinfo{author}{\bibfnamefont{Y.~C.} \bibnamefont{Hu}},
  \bibinfo{author}{\bibfnamefont{D.~S.} \bibnamefont{Cannell}},
  \bibnamefont{and} \bibinfo{author}{\bibfnamefont{G.}~\bibnamefont{Ahlers}},
  \bibinfo{journal}{Rev. Sci. Instrum.} \textbf{\bibinfo{volume}{67}},
  \bibinfo{pages}{2043} (\bibinfo{year}{1996}).

\bibitem[{\citenamefont{Fauve et~al.}(1981)\citenamefont{Fauve, Laroche, and
  Libchaber}}]{fauve:l:l:1981}
\bibinfo{author}{\bibfnamefont{S.}~\bibnamefont{Fauve}},
  \bibinfo{author}{\bibfnamefont{C.}~\bibnamefont{Laroche}}, \bibnamefont{and}
  \bibinfo{author}{\bibfnamefont{A.}~\bibnamefont{Libchaber}},
  \bibinfo{journal}{J. Physique--Lettres} \textbf{\bibinfo{volume}{42}},
  \bibinfo{pages}{455} (\bibinfo{year}{1981}).

\bibitem[{\citenamefont{{Bozhko} and {Putin}}(2003)}]{Bozhko:2003fk}
\bibinfo{author}{\bibfnamefont{A.}~\bibnamefont{{Bozhko}}} \bibnamefont{and}
  \bibinfo{author}{\bibfnamefont{G.}~\bibnamefont{{Putin}}},
  \bibinfo{journal}{Magnetohydrodynamics} \textbf{\bibinfo{volume}{39}},
  \bibinfo{pages}{147} (\bibinfo{year}{2003}).

\bibitem[{\citenamefont{{Huke} and {L{\"u}cke}}(2005)}]{Huke:2005fk}
\bibinfo{author}{\bibfnamefont{B.}~\bibnamefont{{Huke}}} \bibnamefont{and}
  \bibinfo{author}{\bibfnamefont{M.}~\bibnamefont{{L{\"u}cke}}},
  \bibinfo{journal}{Journal of Magnetism and Magnetic Materials}
  \textbf{\bibinfo{volume}{289}}, \bibinfo{pages}{264} (\bibinfo{year}{2005}).

\bibitem[{\citenamefont{{Xu} and {Andereck}}(2003)}]{Xu:2003kx}
\bibinfo{author}{\bibfnamefont{H.}~\bibnamefont{{Xu}}} \bibnamefont{and}
  \bibinfo{author}{\bibfnamefont{C.~D.} \bibnamefont{{Andereck}}},
  \bibinfo{journal}{APS Meeting Abstracts} pp. \bibinfo{pages}{D7+}
  (\bibinfo{year}{2003}).

\bibitem[{\citenamefont{Koschmieder}(1993)}]{koschmieder:1993}
\bibinfo{author}{\bibfnamefont{E.~L.} \bibnamefont{Koschmieder}},
  \emph{\bibinfo{title}{{B}\`{e}nard Cells and Taylor Vortices}}
  (\bibinfo{publisher}{Cambridge University Press}, \bibinfo{year}{1993}).

\bibitem[{\citenamefont{Iida and Guthrie}(1988)}]{iida:g:1988}
\bibinfo{author}{\bibfnamefont{T.}~\bibnamefont{Iida}} \bibnamefont{and}
  \bibinfo{author}{\bibfnamefont{R.~I.~L.} \bibnamefont{Guthrie}},
  \emph{\bibinfo{title}{The Physical Properties of Liquid Metals}}
  (\bibinfo{publisher}{Clarendon Press}, \bibinfo{address}{Oxford},
  \bibinfo{year}{1988}).

\bibitem[{\citenamefont{Kuo}(1999)}]{Kuo:1999fk}
\bibinfo{author}{\bibfnamefont{A.~C.~M.} \bibnamefont{Kuo}},
  \bibinfo{journal}{Polymer Data Handbook} pp. \bibinfo{pages}{411--435}
  (\bibinfo{year}{1999}).

\end{thebibliography}

\clearpage

\begin{figure}[hftbp]
\centering
\caption[Apparatus]{A sectional view of the apparatus design.  See text for label identification.  The ultrasound camera (not shown) sits above the top window ($O$). (color online)} 
\label{fig01} 
\end{figure}

\begin{figure}[hftbp]
\centering
\caption[Transducer]{An exploded view of the ultrasound transducer ($L$) kinematically mounted to three posts ($M$) which are threaded through bushings in the bottom bath flange ($N$).  Also labeled are the jets ($G$) which direct warm water across the bottom surface of the lower sapphire window. (color online)} 
\label{fig02} 
\end{figure}

\begin{figure}[hftbp]
\centering
\caption[Ultrasound camera]{A lens-side up view of the ultrasound camera.  Labeled are the acoustic lenses ($A$), the focus knob ($B$) and the fluid fill port ($C$).  The ruler on the table is six inches long.  (Photo courtesy of Imperium, Inc.; color online)}
\label{fig03} 
\end{figure}

\begin{figure}[hftbp]
\centering
\caption[apparatus photo]{A photograph of the assembled apparatus sitting atop an aluminum platform with a panel for making electrical and plumbing connections. (color online)} 
\label{fig04} 
\end{figure}

\begin{figure}[hftbp]
\centering
\caption[Connections]{External connections between the apparatus and the PC.  These can be divided into three systems.  On the left is the temperature monitoring system; in the middle is the ultrasound imaging system; on the right is the bath regulation system.} 
\label{fig05} 
\end{figure}

\begin{figure}[hftbp]
\centering
\caption[focusing]{Focusing the ultrasound camera on a paperclip and metal washer placed atop the convection cell.}
\label{fig06} 
\end{figure}

\begin{figure}[hftbp]
\centering
\caption[alignment]{Interferometry setup used for alignment of the sapphire windows.}
\label{fig07} 
\end{figure}

\begin{figure}[hftbp]
\centering
\caption[Apparatus]{Thermometer detail.  (color online)} 
\label{fig08} 
\end{figure}

\begin{figure}[hftbp]
\centering
\caption[Temperature]{Top graph:  Time series of bath temperature.  Bottom graph: Power spectral density.}
\label{fig09} 
\end{figure}

\begin{figure}[hftbp]
\centering
\caption[rolls_PDMS]{The onset of straight convection rolls in 5 cSt PDMS with $P\!r = 67.8$.  (a) $\Delta T = 3.22$ degrees and $R\!a = 910$.  (b) $\Delta T = 10.89$ degrees and $R\!a = 3082$.}
\label{fig10} 
\end{figure}

\begin{figure}[hftbp]
\centering
\caption[rolls_EFH1_tilted]{Convective structures above onset in EFH1 ferrofluid with $P\!r = 101.2$.  The cell was tilted by 2.89 degrees.  (a) $\Delta T = 15.44$ degrees and $R\!a = 1986$.  (b) $\Delta T = 26.89$ degrees and $R\!a = 3460$.}
\label{fig11} 
\end{figure}

\clearpage

\begin{table}
\caption{\label{tab:cells}Experimental cells constructed for study of different fluids.  The sample space is circular and has a diameter of approximately 9 cm.  Cell 3 has an optional rectangular insert which converts its aspect ratio to $8\times2$. }
\begin{ruledtabular}
\begin{tabular}{lccr}
Cell & height (cm) & aspect ratio & sidewall material\\
\hline
1 & 0.514 & 17.8 & stainless steel\\
2 & 0.396 & 23.1 & acetal\\
3 & 0.295 & 31.0 (8$\times$2) & acetal \\
4 & 0.226 & 17.8 & acetal \\
5 & 0.340 & 26.8 & acetal \\
\end{tabular}
\end{ruledtabular}
\end{table}

\clearpage

\begin{table}
\caption{\label{tab:hg} Theoretically predicted stability ranges for straight rolls in liquid mercury confined in our cells (1-5), based on the calculations of Busse \citep{busse:1978}.  $\Delta T_{lo}$ is the temperature difference required for $R\!a = 1708$, the lower stability boundary of the Busse balloon; $\Delta T_{hi}$ is the temperature difference required for $R\!a = 1900$, the approximate upper stability boundary of the Busse balloon.  Shorter cells have a larger stability range.}
\begin{ruledtabular}
\begin{tabular}{lcccc}
Cell & Fluid height (cm) & $\Delta T_{lo}$ & $\Delta T_{hi}$ & stability range (C)\\
\hline
1 & 0.514 & 4.82 & 5.37 & 0.55\\
2 & 0.396 & 10.6 & 11.7 & 1.1 \\
3 & 0.295 & 25.5 & 28.4 & 2.9 \\
4 & 0.226 & 56.7 & 63.1 & 6.4 \\
5 & 0.340 & 16.7 & 18.5 & 1.8 \\
\end{tabular}
\end{ruledtabular}
\end{table}

\clearpage

\begin{table*}[hftbp]
\caption[Physical properties of selected materials]{Comparison of the approximate thermo-physical properties of selected materials at 25 degrees Celsius.  Listed are: density ($\rho$), thermal expansion coefficient ($\alpha$), thermal conductivity ($\lambda$), thermal diffusivity ($\kappa$), kinematic viscosity ($\nu$), Prandtl number ($P\!r$), critical temperature difference ($\Delta T_c$), and the vertical and horizontal diffusion times ($\tau_v$ and $\tau_h$).   $\Delta T_c$, $\tau_v$, and $\tau_h$ were computed using 5.19 mm for the fluid depth and 91.4 mm for the cell diameter.}
\label{tab:properties} 
\begin{ruledtabular}
\renewcommand{\arraystretch}{1.5}  
\begin{tabular}{|cc|r|r|r|r|r|}
\multicolumn{2}{|c|}{property} & sapphire & mercury \citep{iida:g:1988} & water & PDMS \citep{Kuo:1999fk} & EFH1 \\ \hline \hline
$\rho$ & $\left(\frac{\mbox{g}}{\mbox{cm}^3}\right)$ & 4.02 & 13.5 & 0.997 & 0.913 & 1.169 \\ \hline
$\alpha$ & $\left(\frac{1}{K}\right)$ & $-0.0000062$ & $-0.00018$ & $-0.00021$ & $-0.001$ & $-0.00075$ \\ \hline
$\lambda$ & $\left(\frac{\mbox{erg}}{\mbox{s}\cdot\mbox{cm}\cdot\mbox{K}}\right)$ & $4,000,000$ & $830,000$ & $61,000$ & $12,000$ & $15,000$ \\ \hline
$\kappa$ & $\left(\frac{\mbox{cm}^2}{\mbox{s}}\right)$ & 1.33 & 0.044 & $0.00146$ & $0.000786$ & $0.000845$ \\ \hline
$\nu$ & $\left(\frac{\mbox{cm}^2}{\mbox{s}}\right)$ & --- & $0.0015$ & $0.0086$ & $0.055$ & $0.086$\\ \hline
$P\!r$ & --- & --- & 0.034 & 5.87 & 69.7 & 101\\ \hline
$\Delta T_c$ & $\left(K\right)$ & --- & 4.82 & 0.76 & 0.53  & 1.24 \\ \hline
$\tau_v$ & $\left(s\right)$ & --- & 5.99 & 181 & 336 & 312\\ \hline
$\tau_h$ & $\left(s\right)$ & --- & 107 & 3220 & 5980 & 5560 \\ \hline
\end{tabular}
\end{ruledtabular}
\end{table*}

\end{document}